\documentclass[aip, amsmath,amssymb, reprint,]{revtex4-1}

\usepackage{graphicx}
\usepackage{dcolumn}
\usepackage{bm}
\usepackage[utf8]{inputenc}
\usepackage[T1]{fontenc}
\usepackage{mathptmx}
\usepackage{etoolbox}
\usepackage{braket}
\usepackage{siunitx}

\makeatletter
\def\@email#1#2{
 \endgroup
 \patchcmd{\titleblock@produce}
  {\frontmatter@RRAPformat}
  {\frontmatter@RRAPformat{\produce@RRAP{*#1\href{mailto:#2}{#2}}}\frontmatter@RRAPformat}
  {}{}
}
\makeatother

\begin{document}

\title[Engineering local strain for single-atom nuclear acoustic resonance in silicon]{Engineering local strain for single-atom nuclear acoustic resonance in silicon}

\author{Laura A. O'Neill}
\affiliation{School of Electrical Engineering \& Telecommunications, UNSW Sydney, NSW 2052, Australia}
\author{Benjamin Joecker}
\affiliation{School of Electrical Engineering \& Telecommunications, UNSW Sydney, NSW 2052, Australia}
\author{Andrew D. Baczewski}
\affiliation{Center for Computing Research, Sandia National Laboratories, Albuquerque, NM, 87123, USA}
\author{Andrea Morello}
\email{a.morello@unsw.edu.au}
\affiliation{School of Electrical Engineering \& Telecommunications, UNSW Sydney, NSW 2052, Australia}

\date{\today}

\begin{abstract}
Mechanical strain plays a key role in the physics and operation of nanoscale semiconductor systems, including quantum dots and single-dopant devices. Here we describe the design of a nanoelectronic device where a single nuclear spin is coherently controlled via nuclear acoustic resonance (NAR) through the local application of dynamical strain. The strain drives spin transitions by modulating the nuclear quadrupole interaction. We adopt an AlN piezoelectric actuator compatible with standard silicon metal-oxide-semiconductor processing, and optimize the device layout to maximize the NAR drive. We predict NAR Rabi frequencies of order 200~Hz for a single $^{123}$Sb nucleus in a wide region of the device. Spin transitions driven directly by electric fields are suppressed in the center of the device, allowing the observation of pure NAR. Using electric field gradient-elastic tensors calculated by density-functional theory, we extend our predictions to other high-spin group-V donors in silicon, and to the isoelectronic $^{73}$Ge atom. 
\end{abstract}

\maketitle

Mechanical strain is a key design parameter for modern solid-state devices, both classical and quantum. In classical microelectronics, strain is used to increase carrier mobility and has been crucial to advancing device miniaturization \cite{thompson2006uniaxial}. Strained heterostructures can confine highly mobile two-dimensional electron gases \cite{schaffler1997high}, used both in classical high-frequency devices and in quantum applications such as quantum dots \cite{hanson2007spins,zwanenburg2013silicon,chatterjee2021semiconductor}, quantum Hall devices \cite{guinea2010energy} and topological insulators \cite{brune2011quantum}. It is well established that local strain strongly affects the properties of gate-defined quantum dots \cite{thorbeck2015formation,park2016electrode,spruijtenburg2018fabrication} and dopants in silicon \cite{dreher2011electroelastic,franke2015interaction,mansir2018linear,pla2018strain}. 

The above examples pertain to static strain. Dynamic strain, and its quantized limit (phonons), constitute instead the ``next frontier'' of hybrid quantum systems \cite{kurizki2015quantum}. Circuit quantum acoustodynamics \cite{manenti2017circuit} aims at hybridizing acoustic excitations with other quantum systems on a chip. Pioneering experiments coupled superconducting qubits to localized acoustic modes of mechanical resonators \cite{oconnell2010quantum} or traveling modes of surface acoustic waves \cite{gustafsson2014propagating}. Proposals exist for hybridizing phonons with the valley-orbit states of donors in silicon \cite{soykal2011sound}. Recent efforts include the coherent drive of spins in solids such as diamond \cite{barfuss2015strong,golter2016optomechanical,lee2017topical} and silicon carbide \cite{whiteley2019spin,maity2020coherent}, and the strong coupling between magnons and phonons \cite{zhang2016cavity}. Phononic quantum networks \cite{habraken2012continuous} can be designed to link acoustically driven quantum systems. 

In this paper, we assess the possibility of controlling the quantum state of a single nuclear spin using dynamic mechanical strain, i.e. the nuclear acoustic resonance (NAR) of a single atom. NAR was observed long ago in bulk antiferromagnets \cite{melcher1968direct} and semiconductors \cite{sundfors1974experimental,sundfors1979nuclear}. It is a very weak effect, and its development has been essentially abandoned after the 1980s. However, the recent demonstration of nuclear electric resonance (NER) in a single $^{123}$Sb nuclear spin in silicon \cite{asaad2020coherent} shows that it is possible to coherently drive a nuclear spin by resonant modulation of the electric field gradient  (EFG) $\mathcal{V}_{\alpha\beta}$ ($\alpha, \beta = x,y,z$) at the nucleus. Here we study the case where the EFG is caused by a time-dependent local strain $\epsilon_{\alpha\beta}$ produced by a piezoelectric actuator. The relation between EFG and strain is described by the gradient-elastic tensor $S$, which was also obtained from the NER experiment in Ref.~\onlinecite{asaad2020coherent}. We expand our analysis by using $S$ values obtained from ab-initio density functional theory (DFT) models, covering the $^{75}$As, $^{123}$Sb and $^{209}$Bi donor nuclei, and the isoelectronic $^{73}$Ge element.

Consider a nuclear spin $I$ with gyromagnetic ratio $\gamma_n$, placed in a static magnetic field $B_0 \parallel z$. For the purpose of this discussion we assume that the nucleus is isolated, i.e. it is not hyperfine- or dipole-coupled to an electron. A coupled electron is necessary during the readout phase\cite{pla2013high}, but can be removed at all other times. The isolated nucleus is described in the basis of the states $\ket{m_{I}}, m_I=-I \ldots I-1, I$ representing the projections of the spin along the $z$-axis, i.e. the eigenvectors of the Zeeman Hamiltonian (in frequency units) 
\begin{align}
    \hat{H}_{\rm Z} = -\gamma_n B_0 \hat{I}_z.
\end{align}
For nuclei with $I>1/2$, a static EFG couples to the electric quadrupole moment $q_{\rm n}$ via the Hamiltonian
\begin{align}
    \hat{H}_{\rm Q}=\frac{e q_{\rm n}}{2I(2I-1)h}\sum_{\alpha,\beta}\mathcal{V}_{\alpha\beta} \hat{I}_\alpha\hat{I}_\beta,
    \label{eq:Hq}
\end{align}
where $e$ is the elementary charge and $h$ is Planck's constant. The quadrupole interaction splits the nuclear resonance frequencies $f_{m_I-1\leftrightarrow m_I}$ between pairs of eigenstates as: 
\begin{align}
    f_{m_I-1\leftrightarrow m_I}=\gamma_n B_0 + (m_I-\frac{1}{2})\frac{e q_n}{2I(2I-1)h}\left(\mathcal{V}_{xx}+\mathcal{V}_{yy}-2\mathcal{V}_{zz}\right)
    \label{eq:spectrum}
\end{align}
and allows addressing individual transitions. Spin transitions can be driven by standard nuclear magnetic resonance (NMR), but also by resonant modulation of the EFG via the off-diagonal Hamiltonian
\begin{align}
    \delta \hat{H}_{\rm Q}=\frac{e q_n}{2I(2I-1)h}\sum_{\alpha,\beta}\delta\mathcal{V}_{\alpha\beta} \hat{I}_\alpha\hat{I}_\beta,
\end{align}
where $\delta\mathcal{V}_{\alpha\beta}$ denotes the amplitude of the time-varying EFG.

For $\Delta m_I=\pm 1$ transitions, the nuclear quadrupolar Rabi frequency $f_{m_{I}-1\leftrightarrow m_{I}}^{\rm Rabi} = |\braket{m_{I}-1|\delta \hat{H}_{\rm Q}|m_{I}}|$ simplifies to
\begin{align}
    f_{m_{I}-1\leftrightarrow m_{I}}^{\rm Rabi}
    = \frac{e |q_n|}{2I(2I-1)h}\alpha_{m_{I}-1\leftrightarrow m_{I}}\left|\delta\mathcal{V}_{xz}+i\delta\mathcal{V}_{yz}\right|,
    \label{Eq:fRabi}
\end{align}
where $\alpha_{m_{I}-1\leftrightarrow m_{I}}$=$|\braket{m_{I}-1| \hat{I}_\beta\hat{I}_z+\hat{I}_z\hat{I}_\beta|m_{I}}|$ for $\beta=x,y$.

In the case of NAR, a time-dependent strain $\delta \epsilon_{\alpha\beta}$ periodically deforms the local charge environment of the nucleus and creates an EFG modulation described by the gradient-electric tensor $S$. This effect depends on the host crystal and its orientation with respect to the coordinate system in which $S$ is defined. For the T$_{\rm d}$ symmetry of a substitutional lattice site in silicon, $S$ is completely defined by two unique elements $S_{11}$ and $S_{44}$. In Voigt's notation and with the Cartesian axes aligned with the $\braket{100}$-crystal axis, e.g. $z\parallel[100]$, $x\parallel[010]$, and $y\parallel[001]$:
\begin{align}
        \begin{pmatrix}
        \delta\mathcal{V}_{xx}\\
        \delta\mathcal{V}_{yy}\\
        \delta\mathcal{V}_{zz}\\
        \delta\mathcal{V}_{yz}\\
        \delta\mathcal{V}_{xz}\\
        \delta\mathcal{V}_{xy}
    \end{pmatrix}
    =
    \begin{pmatrix}
        S_{11} & \frac{-S_{11}}{2} & \frac{-S_{11}}{2} &0&0&0\\
        \frac{-S_{11}}{2} & S_{11} & \frac{-S_{11}}{2} &0&0&0\\
        \frac{-S_{11}}{2} & \frac{-S_{11}}{2} & S_{11} &0&0&0\\
        0&0&0& S_{44} & 0 & 0\\
        0&0&0& 0 & S_{44} & 0\\
        0&0&0& 0 & 0 & S_{44}
    \end{pmatrix}
    \cdot
    \begin{pmatrix}
        \delta\epsilon_{xx}\\
        \delta\epsilon_{yy}\\
        \delta\epsilon_{zz}\\
        2\delta\epsilon_{yz}\\
        2\delta\epsilon_{xz}\\
        2\delta\epsilon_{xy}
    \end{pmatrix},
    \label{eq: Stensor}
\end{align}
where the factor 2 in the shear components arises because the $S$-tensor is defined with respect to engineering strains. Crucially, for a magnetic field $B_0 \parallel z$ aligned with a $\braket{100}$ crystal orientation, Eq.~\ref{Eq:fRabi} and \ref{eq: Stensor} yield the NAR driving frequency
\begin{align}
    f_{m_{I}-1\leftrightarrow m_{I}}^{\rm Rabi,NAR} = \alpha_{m_{I}-1\leftrightarrow m_{I}}\frac{e |q_n|}{2I(2I-1)h} 2 S_{44} \sqrt{{\delta\epsilon_{xz}}^2+{\delta\epsilon_{yz}}^2},
    \label{Eq:fNAR}
\end{align}
which exclusively depends on shear strain components that couple to the EFG via $S_{44}$. Rotating the magnetic field away from the principal crystal axis, e.g. $z\parallel[110]$, would increase the contribution of uniaxial strain components, proportional to $S_{11}$. Since $S_{44}>S_{11}$ in all cases (see Table~\ref{tab:Results}), the strongest acoustic drive is obtained when $B_0 \parallel \braket{100}$. 

A dynamic EFG can also be created by a time-dependent electric field $\delta E_{\alpha}$ which distorts the bond orbitals coordinating the donor. This process, leading to NER\cite{asaad2020coherent}, is described by the $R$-tensor
\begin{align}
    \begin{pmatrix}
        \delta \mathcal{V}_{xx}\\
        \delta \mathcal{V}_{yy}\\
        \delta \mathcal{V}_{zz}\\
        \delta \mathcal{V}_{yz}\\
        \delta \mathcal{V}_{xz}\\
        \delta \mathcal{V}_{xy}
    \end{pmatrix}
    =
    \begin{pmatrix}
        0&0&0\\
        0&0&0\\
        0&0&0\\
        R_{14} & 0 & 0\\
        0 & R_{14} & 0\\
        0 & 0 & R_{14}
    \end{pmatrix}
    \cdot
    \begin{pmatrix}
        \delta E_{x}\\
        \delta E_{y}\\
        \delta E_{z}\\
    \end{pmatrix}.
    \label{eq: Rtensor}
\end{align}
Notably, the resulting NER driving frequency
\begin{align}
    f_{m_{I}-1\leftrightarrow m_{I}}^{\rm Rabi,NER} = \alpha_{m_{I}-1\leftrightarrow m_{I}}\frac{e |q_n|}{2I(2I-1)h} R_{14} \sqrt{{\delta E_{x}}^2+{\delta E_{y}}^2}
    \label{Eq:fNER}
\end{align}
only depends on electric field components perpendicular to $B_0 \parallel z$. In a device where NAR is driven by a piezoelectric actuator, the time-varying strain is necessarily accompanied by a time-varying electric field, but the above observations will allow us to engineer a layout that maximizes NAR while largely suppressing NER.

\begin{figure}[htb!]
	\centering
	\includegraphics[width=1\columnwidth]{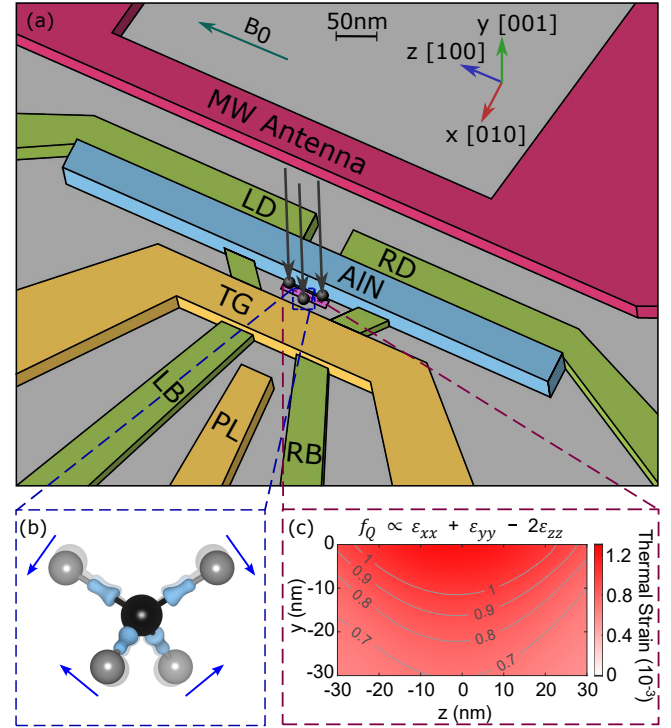}
	\caption{(a) Device geometry for nuclear acoustic resonance, based upon standard donor qubit devices but modified to include a 55~nm thick piezoelectric actuator (AlN, blue). A single-electron transistor is formed by an electron gas induced by the top gate (TG, yellow), controlled by the plunger gate (PL, yellow), left and right barriers (LB, RB, green). Left and right donor gates (LD, RD, green) control the donor electrochemical potential. The piezoactuator creates a time-dependent strain when applying a radiofrequency voltage $V_{\rm RF}\cos(2\pi f_{m_I-1\leftrightarrow m_I}t)$ to LB, LD, and $-V_{\rm RF}\cos(2\pi f_{m_I-1\leftrightarrow m_I}t)$ to RB, RD. A microwave antenna (magenta) is used to induce magnetic resonance transitions as necessary for nuclear spin readout via an electron spin ancilla. A static magnetic field $B_0$ is assumed applied along the $z\equiv [100]$ axis. The design assumes the centre of the 60(W)x30(H)x10(D)~nm$^3$ implantation window is located 30~nm from the top gate TG. (b) Sketch (generated using VESTA~\cite{momma2011vesta}) of strain-induced atomic bond distortion for a substitutional donor (black) in silicon (grey). (c) Distribution of static strain in the device, caused by differential thermal expansion. We plot the components $\epsilon_{xx}+\epsilon_{yy}-2\epsilon_{zz}$ responsible for the nuclear quadrupole splitting $f_{\rm Q}$ (Eq.~\ref{eq:fQ}).}
	\label{Fig:1}
\end{figure}
We thus propose the device structure shown in Fig.~\ref{Fig:1}. It is similar to the standard layout adopted in metal-oxide-semiconductor (MOS) compatible single-donor devices in silicon \cite{morello2009architecture,morello2020donor}, including a single-electron transistor (SET) for electron spin readout via spin-to-charge conversion \cite{morello2010single}, an on-chip microwave antenna \cite{dehollain2012nanoscale} to drive electron \cite{pla2012single} and nuclear \cite{pla2013high} spin resonance transitions, and electrostatic gates to locally control the potential in the device. The same gates, connected to control lines with $\sim 100$~MHz bandwidth, can be used to deliver oscillating electric fields \cite{asaad2020coherent}. A group-V donor or isoelectronic center with nuclear spin $I>1/2$ is introduced by ion implantation. To address an isoelectronic center like $^{73}$Ge, the structure should further include a lithographically-defined quantum dot\cite{hensen2020silicon} to host an additional electron, hyperfine-coupled to the nucleus, as recently demonstrated with $^{29}$Si.

We introduce two changes to the standard layout. First, we include a strip of piezoelectric material, placed on top of the implantation region between the gates and the SET, to create a time-dependent local strain $\delta \epsilon_{\alpha\beta}$ upon application of an oscillating voltage $V_{\rm RF}$ to the gates. Second, we align the piezoelectric and the gates with the $[100]$ crystal direction, along which a static external magnetic field $B_0 \sim 1$~T is applied ($z$-axis). This requires rotating the device layout by $45^{\circ}$ compared to standard donor devices, where $B_0$ and gates are aligned along [110] \cite{tenberg2019electron}, which is the natural cleaving face for silicon wafers. 

We model the device geometry in the modular COMSOL multiphysics software. A $\SI{2}{\micro\meter} \times \SI{2}{\micro\meter} \times\SI{2}{\micro\meter}$ silicon substrate is capped by an \SI{8}{\nano\meter} thick SiO$_2$ layer. The aluminum gates, covered by \SI{2}{\nano\meter} of Al$_2$O$_3$ through oxidation, and the piezoelectric actuator are placed on top. 
We use the `AC/DC Module' to compute the electrostatics, the `Structural Mechanics Module' for thermal deformation, and combined multiphysics simulations for the piezoelectric coupling.
The static strain, created upon cooling the device from 850~\textdegree{}C to 0.2~K in two stages by the difference in thermal expansion coefficients among different materials in the stack, is modelled as described in Ref.~\onlinecite{asaad2020coherent}. Fig.~\ref{Fig:1}c shows the components of the static strain that cause the splitting $f_{\rm Q}$ between nuclear resonance frequencies in Eq.~\ref{eq:spectrum}:
\begin{align}
    f_{\rm Q}= \frac{e q_n}{2I(2I-1)h}\frac{3}{2}S_{11}\left(\epsilon_{xx}+\epsilon_{yy}-2\epsilon_{zz}\right).
    \label{eq:fQ}
\end{align}
In the center of the implantation region, near the Si/SiO$_2$ interface, we predict $|f_{\rm Q}|=\SI{14}{\kilo\hertz}$ for the $^{123}$Sb nucleus (see Table~\ref{tab:Results} for other nuclei), ensuring that the resonance lines are well resolved. In the the electrostatic simulations, the idle gate voltages are set to $V_{\rm LB}=\SI{0}{\volt}$, $V_{\rm RB}=\SI{0}{\volt}$, $V_{\rm PL}=\SI{0}{\volt}$, $V_{\rm TG}=\SI{1.8}{\volt}$, $V_{\rm LD}=\SI{0}{\volt}$, $V_{\rm RD}=\SI{0}{\volt}$, and $V_{\rm MW}=\SI{0}{\volt}$. Additionally, we ground the Si/SiO$_2$ interface under the SET to model the effect of the conducting electron channel \cite{morello2009architecture,morello2010single}.
The COMSOL material library conveniently provides all other parameters.

We choose aluminum nitride (AlN) as the piezoelectric actuator. Although other materials such as ZnO and PZT (Pb[Zr$_x$Ti$_{1-x}$]O$_3$) have stronger piezoelectric response, AlN has the key advantage of being compatible with the MOS fabrication flow. Other piezoelectrics contain fast-diffusing elements which would contaminate the device and potentially the process tools.

\begin{figure}[t]
	\centering
	\includegraphics[width=1\columnwidth]{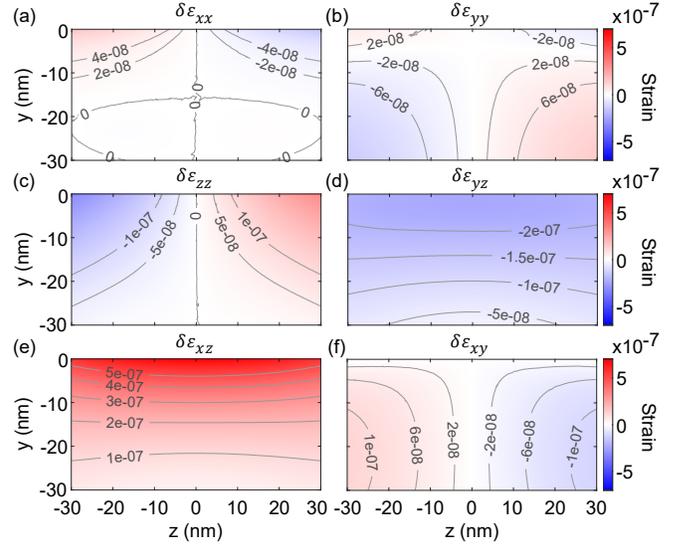}
	\caption{Amplitudes of the periodic strain variation in the implantation window during the acoustic drive.  The uniaxial strain components (a) $\delta\epsilon_{xx}$, (b) $\delta\epsilon_{yy}$ and (c) $\delta\epsilon_{zz}$ and shear strain components (d) $\delta\epsilon_{yz}$, (e) $\delta\epsilon_{xz}$ and (f) $\delta\epsilon_{xy}$ were calculated using the difference in strain between static gate voltages $V_{LD}=V_{LB}=V_{RD}=V_{RB}=0$~V and peak driving amplitudes $V_{LD}=V_{LB}=100$~mV and $V_{RD}=V_{RB}=-100$~mV.  Shown are cross sections below the Si/SiO$_2$ interface in the centre of the implantation window, located 30~nm from the SET top gate, as indicated in Fig.~\ref{Fig:1}.  Shear components $\delta\epsilon_{yz}$ and $\delta\epsilon_{xz}$ are the largest, indicating strongest acoustic drive when $B_0\parallel [100]$ axis.  }
	\label{Fig:2}
\end{figure}

Figure~\ref{Fig:2} shows the maps of dynamical strain $\delta \epsilon_{\alpha\beta}$ along a vertical cross-section of the device, assuming that $V_{\rm RF}$ has opposite phase on the left and right gates, and 100~mV peak amplitude. The model clearly shows that the shear strain $\delta\epsilon_{yz}$ and $\delta\epsilon_{xz}$ is the dominant component in the center of the device, as required for fast acoustic drive as per Eq.~\ref{Eq:fNAR}.

\begin{figure}[htb!]
	\centering
	\includegraphics[width=1\columnwidth]{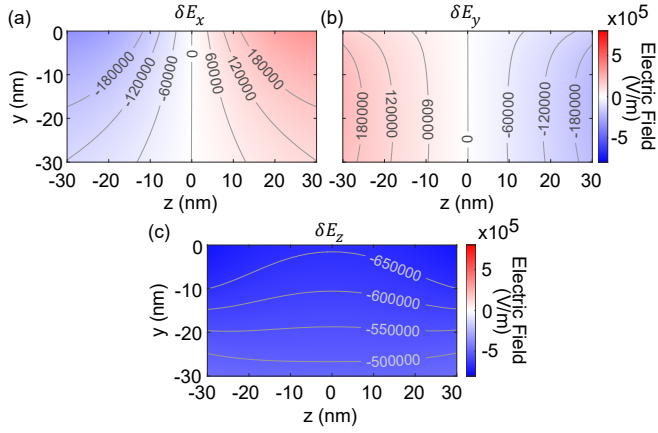}
	\caption{Amplitudes of the electric field variation in the implantation window during the acoustic drive.  The electric field components (a) $\delta E_{x}$, (b) $\delta E_{y}$ and (c) $\delta E_{z}$ were calculated using the difference in electric field between static gate voltages $V_{LD}=V_{LB}=V_{RD}=V_{RB}=0$~V and peak driving amplitudes $V_{LD}=V_{LB}=100$~mV and $V_{RD}=V_{RB}=-100$~mV.  Shown is the same cross section as in Fig.~\ref{Fig:1}. For $B_0 \parallel [100]$ axis, the electric drive solely depends on  $\delta E_{x}$ and $\delta E_{y}$ (see Eq.~\ref{Eq:fNER}), which both vanish at the center of the device.}
	\label{Fig:3}
\end{figure}

To assess the strength of the electric contribution to the nuclear drive, we use COMSOL to model the amplitude of the electric field change $\delta E_{\alpha}$ produced by $V_{\rm RF}$, plotted in Fig.~\ref{Fig:3}. Our chosen device layout, having mirror symmetry around the $z=0$ plane, and the applied $V_{\rm RF}$ having opposite phase on the left and right gates, make $\delta E_x$ and $\delta E_y$ vanish in the center of the device. 

\begin{figure}[htb!]
	\centering
	\includegraphics[width=1\columnwidth]{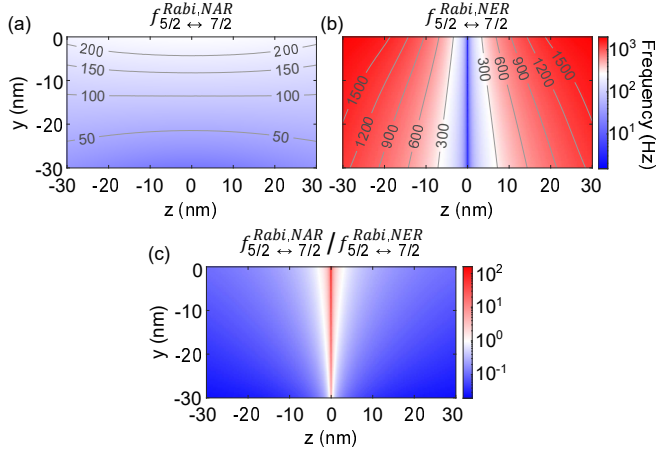}
	\caption{The nuclear acoustic resonance (NAR) and nuclear electric resonance (NER) Rabi frequencies were calculated for the $^{123}$Sb $\ket{5/2}\leftrightarrow\ket{7/2}$ transition with the B field oriented along the $z$-axis ($[100]$ crystal axis) using Eq.~\ref{Eq:fNAR} and \ref{Eq:fNER} respectively with $R_{14}=1.7\times10^{12}$~m$^{-1}$ and $S_{44}= 5.9\times10^{22}$~V/m$^2$.  (a) The NAR transition frequencies are uniformly distributed along the top region of the implantation window with a maximum of around 274~Hz.  (b) The NER transition frequencies are minimal in the center of the implantation window (minimum of around 1.5~Hz).  (c) Their ratio $f_{5/2\leftrightarrow 7/2}^{\rm Rabi,NAR}/f_{5/2\leftrightarrow 7/2}^{\rm Rabi,NER}$ demonstrates the region in which the NAR frequencies are greater than or comparable to the corresponding NER frequencies (maximum ratio of around 160).  A donor in the centre of the implantation window  ($z=0$~nm) at a depth of $y=-5$~nm achieves $f_{5/2\leftrightarrow 7/2}^{\rm Rabi,NAR}=190$~Hz while keeping $f_{5/2\leftrightarrow 7/2}^{\rm Rabi,NER}=2.7$Hz, corresponding to a ratio of $f_{5/2\leftrightarrow 7/2}^{\rm Rabi,NAR}f_{5/2\leftrightarrow 7/2}^{\rm Rabi,NER}\approx 70$.}
	\label{Fig:4}
\end{figure}

The main result of our work is shown in Fig.~\ref{Fig:4}. We calculate the nuclear Rabi frequencies predicted on the basis of both NAR ($f^{\rm NAR}$, Eq.~\ref{Eq:fNAR}) and NER ($f^{\rm NER}$, Eq.~\ref{Eq:fNER}), using the parameters pertaining the $\ket{5/2}\leftrightarrow\ket{7/2}$ transition of a $^{123}$Sb nucleus \cite{asaad2020coherent}. We find $f^{\rm NAR} \approx 200$~Hz in a wide region of the device, at the shallow depths ($\approx 5-10$~nm) expected for donors implanted at $\sim 10$~keV energy \cite{van2015single,jakob2020deterministic}. For an ionized donor nuclear spin in isotopically enriched $^{28}$Si, where the dephasing time is $T_{\rm 2n}^* \sim 0.1$~s, this value of $f^{\rm NAR}$ is sufficient to ensure high-quality coherent control. 

Consistent with earlier experimental results \cite{asaad2020coherent}, we predict NER Rabi frequencies up to $f^{\rm NER} \approx 1.5$~kHz. However, our design ensures that $f^{\rm NER}$ vanishes in the center of the device. This results in a $\approx 10$~nm wide region where $f^{\rm NAR} \gg f^{\rm NER}$ [Fig.~\ref{Fig:4}(c)], i.e. wherein pure NAR can be observed.

A side effect of the application of strain is the local modulation of the host semiconductor's band structure, which can shift the electrochemical potential of the donor with respect to the SET. This must be minimized to ensure that the charge state of the donor does not change during the NAR drive. The effect of strain on the conduction band can be described via deformation potentials\cite{wilson1961electron}. The dominant contribution is uniaxial strain that shifts the respective valleys by $\delta E^{\rm CB}_{\pm \alpha}=\Xi_{\rm u} \delta\epsilon_{\alpha\alpha}$, where $\Xi_{\rm u}=10.5$~eV \cite{fischetti1996band} for silicon. We estimate a worst-case shift $\delta E^{\rm CB}_{\rm SET}= 0.525~\mu$eV at the SET, and $\delta E_{\rm Donor}= 3.36~\mu$eV at the donor location. These values are orders of magnitude smaller than the electron confinement energies and the Zeeman splitting (the relevant scale for spin readout\cite{morello2010single}), and small enough to be cancelled by compensating voltages on the local gates, if required.

The calculations applied above to $^{123}$Sb can be extended to any other $I>1/2$ nucleus that can be individually addressed in silicon, by simply adapting the values of $S_{11}$ and $S_{44}$. Table.~\ref{tab:Results} presents values calculated using the projector-augmented wave formalism implemented in the Vienna Ab initio Simulation Package (VASP)~\cite{kresse1996efficiency,kresse1996efficient,kresse1999ultrasoft}.
For each dopant species, the EFG at the relevant nucleus is calculated using a supercell of 512 atoms with one singly ionized dopant and a plane-wave cutoff of 500 eV~\cite{petrilli1998electric}.
Having previously established a linear relationship between the EFG and strain up to 1\% for $^{123}$Sb~\cite{asaad2020coherent}, we carry out all EFG calculations for 1\% strain and determine the tensor components from Eq.~\ref{eq: Stensor}.
The numbers in Tab.~\ref{tab:Results} were computed using the SCAN exchange-correlation functional~\cite{sun2015strongly}.
Using other exchange-correlation functionals, LDA~\cite{ceperley1980ground} and PBE~\cite{perdew1996generalized}, leads to a 2-10\% variation in $S_{11}$ and $S_{44}$ with no consistent trends among the species or functionals.
As SCAN best reproduces the bulk elastic properties among the functionals considered, we consider those numbers to be the most reliable and have reported them.
\begin{table}[htb!]
    \centering
    \includegraphics[width=1\columnwidth]{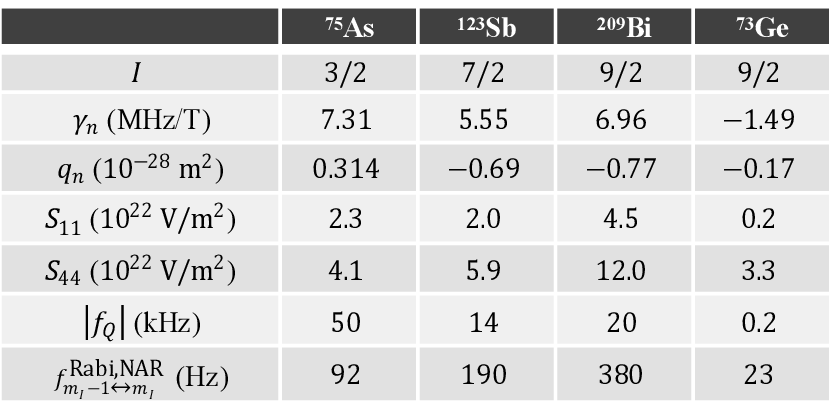}
\caption{Parameter and results for different donors with nuclear spin $I>1/2$. The nuclear gyromagnetic ratios and quadrupole moments are extracted from Ref.~\onlinecite{stone2005table}, where a range of values for $q_{\rm n}$ are reported. The uniaxial $S_{11}$ and shear $S_{44}$ components of the gradient elastic tensor (see Eq.~\ref{eq: Stensor}) were calculated using DFT. The resulting quadrupole splitting $f_{\rm Q}$ (Eq.~\ref{eq:fQ}) is given for a donor located in the centre of the implantation region at depth $y=-5$~nm. The corresponding NAR Rabi frequencies (Eq.~\ref{Eq:fNAR}) are reported for the $m_I=I-1 \leftrightarrow I$ transition.}
\label{tab:Results}
\end{table}

In conclusion, our results show that a simple AlN piezoelectric actuator placed within a standard MOS-compatible donor qubit device is capable of driving coherent NAR transitions in a high-spin group-V donor in silicon. The choice of device layout and magnetic field orientation with respect to the Si crystal axes allows to suppress NER in the center of the device.

The experimental realization of this architecture will provide unique insights into the microscopic interplay between strain and spin qubits in silicon. The exceptional intrinsic spin coherence of nuclear spins in silicon, which results in resonance linewidths $< 10$~Hz, translates into an equivalent spectroscopic resolution in the static (via $f_{\rm Q}$) and dynamic (via $f^{\rm Rabi,NAR}$) strain, detected by an atomic-scale probe. This information can be further correlated to other properties of the spin qubits hosted in the device, such as spin relaxation times \cite{tenberg2019electron}, hyperfine couplings \cite{dreher2011electroelastic,mansir2018linear,pla2018strain} or exchange interactions \cite{voisin2020valley,chan2021exchange,madzik2021conditional}. Furthermore, the mechanical drive of a nuclear spin in an engineered silicon device will inform the prospect of coherently coupling nuclear spins to the quantized motion of high-quality mechanical resonators\cite{ghaffari2013quantum,safavi2019controlling}, realizing a novel form of hybrid quantum system\cite{kurizki2015quantum}.

\begin{acknowledgments}
We thank A. Michael, V. Mourik and A. Saraiva for useful discussions. The research was funded by the Australian Research Council Discovery Projects (Grants No. DP180100969 and DP210103769), the US Army Research Office (Contract no. W911NF-17-1-0200), and the Australian Department of Industry, Innovation and Science (Grant No. AUSMURI000002). A.D.B. was supported by the U.S. Department of Energy, Office of Science, National Quantum Information Science Research Centers (Quantum Systems Accelerator) and Sandia National Laboratories' Laboratory Directed Research and Development program (Project 213048).
Sandia National Laboratories is a multi-missions laboratory managed and operated by National Technology and Engineering Solutions of Sandia, LLC, a wholly owned subsidiary of Honeywell International Inc., for DOE's National Nuclear Security Administration under contract DE-NA0003525. 
The views expressed in this manuscript do not necessarily represent the views of the U.S. Department of Energy or the U.S. Government.
\end{acknowledgments}

\section*{Data Availability Statement}

The data that support the reported findings are available in FigShare at \url{https://doi.org/10.6084/m9.figshare.16529208.v1}.
\section*{References}
%\bibliography{citation}
%merlin.mbs aipnum4-1.bst 2010-07-25 4.21a (PWD, AO, DPC) hacked
%Control: key (0)
%Control: author (8) initials jnrlst
%Control: editor formatted (1) identically to author
%Control: production of article title (0) allowed
%Control: page (1) range
%Control: year (1) truncated
%Control: production of eprint (0) enabled
%

\end{document}